\documentclass[aps,floats,showpacs,superscriptaddress,twocolumn]{revtex4}
\usepackage{tabularx,graphicx}
\usepackage{epsfig}
\usepackage{amsmath}
\usepackage{bbm}
\usepackage{graphicx}

\begin{document}

\title{Localization of electronic states by fullerene charges
in carbon nanotubes}
\author{J. Gonz\'alez}
\affiliation{
        Instituto de Estructura de la Materia.
        Consejo Superior de Investigaciones Cient{\'\i}ficas.
        Serrano 123, 28006 Madrid. Spain.}
\author{F. Guinea}
\affiliation{
        Instituto de Ciencia de Materiales de Madrid.
        Consejo Superior de Investigaciones Cient{\'\i}ficas.
        Cantoblanco. 28049 Madrid. Spain.}

\date{\today}

\begin{abstract}
We study the effects of the electrostatic interaction
produced by charged fullerenes encapsulated in carbon
nanotubes, showing that they are able to modify locally the
electronic density of states in the hybrid system.
In the cases where the interaction is felt as
an attractive potential by the electrons in the
nanotube, localized electronic states are formed in the
nanotubes around the position of the fullerenes.
This produces an effective narrowing of the
gap in semiconducting nanotubes over a distance of a
few nanometers, in agreement with the spatial modulation
of the gap observed in the experiments.

\end{abstract}

\maketitle



Carbon nanotubes have nowadays a great potential for
technological applications in devices at the nanometer
scale. This makes very important the precise knowledge
of their electronic properties, once the first step of
discerning their metallic or insulating behavior has been
completed after confirmation of the theoretical
analyses\cite{saito} by experimental results\cite{wild}.
We have learned from these studies that the low-energy
spectrum of a carbon nanotube may be gapless or not,
depending on the helicity of the tubular structure.

Hybrid structures have been recently synthesized by encapsulation
of fullerenes in the hollow space of the nanotubes. It has been
shown that the insertion of the fullerene molecules in these
so-called peapods leads to a significant modification of the
electronic spectrum of the carbon nanotubes, opening the
possibility to tailor different structures with the desired
functionality. In the experiment reported in Ref.
\onlinecite{horn}, chains of $C_{60}$ molecules have been
encapsulated in the interior of semiconducting nanotubes, showing
that the measured spectra can be explained from the hybridization
of the electronic states in the nanotubes with those in the
fullerenes. In Ref. \onlinecite{lee}, a different kind of peapod
has been studied, formed by the encapsulation of sparse
metallofullerenes $Gd@C_{82}$ in semiconducting nanotubes. In that
case, a striking observation has been the spatial modulation of
the nanotube bandgap, which is narrowed from 0.43 eV down to 0.17
eV at the fullerene positions.

Several electronic structure calculations have shown that some
levels of the fullerene molecular orbitals may lie within the gap
of the semiconducting nanotubes\cite{ihm,oshi}. These results have
been used to explain the narrowing of the bandgap observed
experimentally by claiming that, at the fullerene sites, the
measured gap should correspond to the distance from the top of the
valence band to the lowest unoccupied molecular level within the
nanotube gap. This kind of effective reduction of the gap seems to
be in agreement with the value measured in the experiments but, on
the other hand, it leaves unaccounted a most significant property
of its spatial modulation. Thus, while the effect of the
unoccupied molecular orbital should decay over a distance of about
$3 {\rm \AA }$ along the nanotube, the spatial modulation measured
in Ref. \onlinecite{lee} is much less abrupt, with the narrowing
of the gap taking place over an extension of about 2 or 3 nm.

The distinctive feature of the peapods displaying
the modulation of the bandgap lies in the charge of the
encapsulated fullerenes. For the nanotubes with arrays
of $C_{60}$ molecules inside reported in Ref.
\onlinecite{horn}, no sign of reduction of the
gap has been observed. On the other hand, it is natural
to expect that the Coulomb interaction due to charged
fullerenes may lead to significant effects in the
electronic spectra. In that respect, it has been shown
that, in the case of the peapods with the encapsulated
metallofullerenes, there should be a charge transfer of
about one electron from the latter to the surrounding
nanotube. Thus, the metallofullerenes seem
to have a net positive charge in the samples reported in
Ref. \onlinecite{lee}, making pertinent the discussion
of the Coulomb interaction.

In the present paper, we show that the electrostatic
interaction produced by charged fullerenes is able
to modify locally the electronic density of states of
the peapods. In the cases where the interaction is felt
as an attractive potential by the electrons in the
nanotube, localized electronic states are formed in the
nanotubes around the position of the fullerenes. Their
spatial extension as well as their number can be
estimated by scaling arguments following from the
particular form of the potential. For that purpose,
we can rely on a one-dimensional (1D) low-energy model
of the semiconducting nanotubes, which will allow also
the determination of the bound states formed by the
Coulomb potential within the gap.

We illustrate our 1D low-energy description in the
case of the zig-zag nanotubes. These have a
unit cell with length $3a$, in terms of the $C-C$
distance $a$ . For each transverse array of atoms
$n = 1, 2, \ldots N$ at a given nanotube section $x$,
it is convenient to introduce the Fourier transform of
the electron operators $c (x,n)$
\begin{equation}
c (x,n) \sim \sum_p d_p (x) e^{i 2\pi np/N}
\end{equation}
where the index $p$ labels the different subbands
$p = 0, \pm 1, \ldots \pm N/2$. Within the tight-binding
approach, the hamiltonian for the zig-zag nanotube is
\begin{eqnarray}
 H_{tb}  &  =  &     - t_{\parallel}
           \sum_{p,l} d^+_p (3la) d_p (3la + a)
                                       \nonumber   \\
      &    &  - t_{\perp}  \sum_{p,l}  z_p
       d^+_p (3la + a) d_p (3la + 3a/2)
                                \nonumber   \\
  &    &  - t_{\parallel}  \sum_{p,l}
       d^+_p (3la + 3a/2)  d_p (3la + 5a/2)
                                 \nonumber      \\
  &   &   - t_{\perp}   \sum_{p,l}  z^{*}_p
     d^+_p (3la + 5a/2) d_p (3(l+1)a)
\label{tb}
\end{eqnarray}
with $z_p = 1 +  \exp (i 2\pi p /N  )$ and $l \in Z$. Henceforth
we will consider that the transfer integrals for the two different
bond orientations can be taken approximately equal, $t_{\perp} =
t_{\parallel} = t$. Within this description, each band is obtained
from the diagonalization  of a one dimensional chain with two
orbitals per unit cell, with gap $2 \Delta_p = 2 t | 1 - 2 \cos (
2 \pi p / N ) |$. This simple one-particle description accounts
for the semiconducting character of the zig-zag nanotubes when $N$
is not a multiple of 3.

In the semiconducting nanotube, we can write $N = 3 N' \pm 1$. The
subbands closest to the Fermi level have $p \approx N'$, and a
dispersion typical of a Dirac spinor with Fermi velocity $v_F =
3ta/2$ and a mass parameter $m = \Delta_p$, $\varepsilon (k)
\approx \pm \sqrt{v^2_F k^2 + m^2}$ \cite{kane}. In the following
we assume that the fullerene potential has radial symmetry, and
the subband index $p$ remains a good quantum number. Then, the
calculation of the perturbed energy levels can be factorized into
a set of equivalent problems, one for each subband of the
nanotube. For the sake of describing the many-body effects in the
interaction with the fullerene charge $Q_z$, it is most convenient
to write the model for the low-energy electronic states in the
continuum, in terms of the Dirac spinor field with components
$\Psi_L$ and $\Psi_R$. We will assume that the fullerene charge
has a finite extension as seen along the longitudinal dimension of
the nanotube, over a distance $R$ of the order of the fullerene
radius. The full hamiltonian of the interacting theory becomes
then
\begin{eqnarray}
H  & = &  \int dx ( -i v_F \Psi^+_L \partial_x \Psi_L
      + i v_F  \Psi^+_R \partial_x \Psi_R   \nonumber   \\
& &      \;\;\;\;\;\;\;\;\;\;\;\;\;\;\;\;
  + m \Psi^+_L \Psi_R + m \Psi^+_R \Psi_L ) \nonumber  \\
  &   & + \frac{e^2}{2 } \int dx dy
       (\rho_L (x) + \rho_R (x) )  \frac{1}{|x - y|}
               (\rho_L (y) + \rho_R (y) )     \nonumber    \\
  &   &  -  Q_z e^2 \int dx dy
 \frac{1}{\sqrt{\pi } R}  e^{-(x-z)^2/R^2}  \nonumber  \\
 & & \times \frac{1}{|x - y|}   (\rho_L (y) + \rho_R (y) )
\label{ham}
\end{eqnarray}
where $\rho_L (x), \rho_R (x)$ are the
respective electron density operators for the two
spinor components\cite{neg}.

The effect of the fullerene charge in the electronic
spectrum can be understood by applying the semiclassical
approximation to the electron wavefunctions.
In this approach, we assume that the only significant
effect of the electron-electron interaction within the
nanotube is to induce some charge distribution
screening the net charge of the fullerene.
The screening is not complete, as it is
produced by a system with a gap at the Fermi level, and
it can be encoded at a later stage in the form of a
suitable dielectric function. At this point, we
concentrate therefore on the resolution of the eigenvalue
problem
\begin{eqnarray}
( -i v_F \partial_x + V(x-z) ) \Psi_L + m \Psi_R  & = &
                \varepsilon  \Psi_L      \label{eig1}   \\
( i v_F \partial_x + V(x-z) ) \Psi_R + m \Psi_L  & = &
                \varepsilon  \Psi_R
\label{eig2}
\end{eqnarray}

The system (\ref{eig1})-(\ref{eig2}) can be reduced to
an equation for any of the two components of the Dirac
spinor, and the WKB solutions can be obtained in the
usual fashion. In the semiclassical approximation, the
$\Psi_R $ component reads for instance
\begin{eqnarray}
\Psi_R (x) &  \approx  &   \frac{
  e^{  \pm \frac{1}{2} \int^x dx  \partial_x V(x-z) /
   \sqrt{(\varepsilon - V(x-z))^2 - m^2}  }   }
      { ( (\varepsilon - V(x-z))^2 - m^2 )^{1/4} }
                                        \nonumber   \\
  &  &   \times     e^{  \pm i
   \int^x dx (1/v_F)\sqrt{(\varepsilon - V(x-z))^2 - m^2} }
\end{eqnarray}
The turning points are given by the
roots of the equation $(\varepsilon - V(r_0))^2 = m^2$.
When such points exist, the semiclassical approximation
gives rise to the quantization condition
\begin{equation}
\int_{-r_0}^{r_0} dx \sqrt{(\varepsilon - V(x))^2 - m^2}
   =  v_F \pi \left(n + \frac{1}{2}\right)
\label{qc}
\end{equation}

When the potential $V(x)$ has the long-distance behavior
$V(x) \sim -1/|x|$, it can be shown from Eq. (\ref{qc})
that there is an infinite series of bound
states with energy below and approaching the value of
$m$. If we define
$m - \varepsilon = \epsilon$, the quantization condition
translates into
\begin{equation}
\int_{-r_0}^{r_0} dx \sqrt{( - V(x) - \epsilon )
         ( 2m - V(x) - \epsilon ) }
   =  v_F  \pi \left(n + \frac{1}{2}\right)
\end{equation}
As the parameter $\epsilon $ goes to zero, we observe
that the integral at the left-hand-side of the equation
scales as $1/\sqrt{\epsilon }$, reflecting the divergent
range of integration. Thus, for large values of $n$, we
find the quantization rule
\begin{equation}
m - \varepsilon_n  \sim \frac{1}{(n+1/2)^2}
\end{equation}

The above arguments convey that the Coulomb potential
of the fullerene charge has to induce localized
electronic states within the nanotube gap. We show next
that this conclusion holds after incorporating the
screening of the potential by the electronic charge in
the nanotube. Our continuum model allows actually an
exact treatment of such effect by means
of bosonization methods\cite{bos}. Following these, the
electron density for a given subband near the Fermi level
can be mapped into a boson field $\Phi (x) $ according
to the relation $\frac{1}{2\pi }
\partial \Phi (x) = \rho_L (x) + \rho_R (x)$. In terms
of the boson fields, the hamiltonian (\ref{ham}) can be
expressed in the form
\begin{eqnarray}
H  & = & \frac{1}{2} \int dx  ( 4\pi v_F  \Pi^2 (x)  +
    \frac{v_F}{4\pi }  ( \partial_x \Phi (x) )^2
 +  \frac{2m}{\pi \alpha } \cos ( \Phi (x) )   )
                                        \nonumber    \\
  &   & + \frac{e^2}{8 \pi^2 } \int dx dy
    \partial_x  \Phi (x)  \frac{1}{|x - y|}
           \partial_y \Phi (y)          \nonumber    \\
  &   &  -  \frac{Q_z e^2}{2\pi } \int dx dy
 \frac{1}{\sqrt{\pi } R}  e^{-(x-z)^2/R^2}
      \frac{1}{|x - y|}     \partial_y \Phi (y)
\label{hamb}
\end{eqnarray}
where $\Pi (x) $ is the momentum conjugate of the
field $\Phi (x)$ and $\alpha $ is a short-distance cutoff
needed to regulate the product of fields at the same
point.

\begin{figure}
\begin{center}
\mbox{\epsfxsize 7.5cm \epsfbox{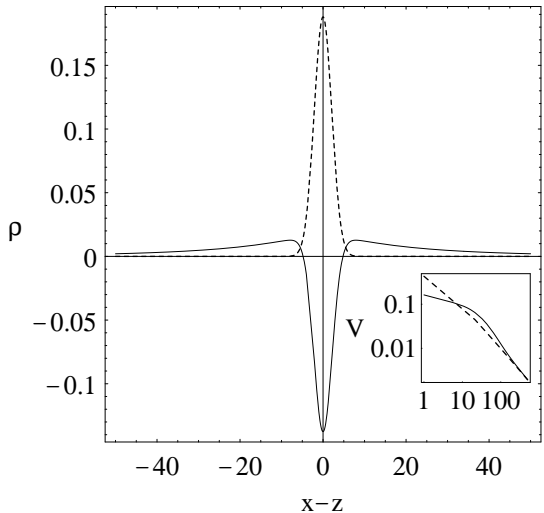}}
\end{center}
\caption{Plot of the model fullerene charge distribution
for $Q_z = 1$ (dashed line) and of the
screening charge distribution induced in the carbon
nanotube (full line). The variable $x$ is measured in
units of the $C$-$C$ distance in the nanotube.
Inset: Log-log plot of the bare Coulomb potential
(dashed line) and of the dressed potential obtained after
considering the screening charge in the nanotube (full
line).}
\label{one}
\end{figure}

From the representation (\ref{hamb}), we observe that the presence
of the charge $Q_z$ induces a nonvanishing average value of the
field $\partial_x \Phi (x)$. As we are interested in low-energy
screening properties, we can expand the $\cos ( \Phi (x) )$
dependence to deal with a hamiltonian quadratic in the boson
fields. The screening charge distribution can be obtained then by
shifting the field $\Phi (x) $ by a suitable function, $\Phi (x) -
f(x) = \tilde{\Phi} (x)$, so that the redefined density has a
vanishing average, $\langle \partial_x \tilde{\Phi} (x) \rangle =
0$. The approach allows an straightforward generalization to the
case of $N_f$ different electron flavors, accounting for spin and
subband degeneracies. Thus, the expression for the screening
electron density induced in the nanotube is in general
\begin{eqnarray}
\rho_s (x)  & = &  \frac{1}{2\pi }
        \langle \partial \Phi (x) \rangle
 =   Q_z  \int \frac{dk}{2 \pi } \cos (k(x-z))
      e^{-R^2 k^2 /4}                 \nonumber        \\
 &   &   \times
    \frac{ \frac{2N_f e^2}{\pi v_F} k^2 \log (1 + k_0/k) }
      {k^2 + \frac{4 m }{\alpha v_F} +
       \frac{2N_f e^2}{\pi v_F} k^2 \log (1 + k_0/k)   }
\end{eqnarray}
where the logarithmic dependences come from the
representation of the bare Coulomb potential in momentum
space\cite{sarma}.
The shape of the screening charge distribution has been plotted in
Fig. \ref{one} for $m/t = 0.1$ and $R = 3a$, taking $N_f = 4$ and
$e^2/v_F \approx 2.7$ . We observe that in general there is only a
partial screening of the fullerene charge $Q_z$, due to the
absence of gapless excitations in the semiconducting nanotube. The
results are not qualitatively changed if other subbands with
higher gaps are also included.

The dressed Coulomb potential $V_d (x-z) $ felt in
the nanotube is created by the sum of the background
charge $Q_z e$ and the screening charge $-e \rho_s (x)$.
Thus we have
\begin{eqnarray}
V_d (x-z) & = & - Q_z e^2 \int \frac{dk}{2 \pi } \cos (k(x-z))
           e^{-R^2 k^2 /4}            \nonumber        \\
 &  &    \times
 \frac{ (k^2 + \frac{4 m }{\alpha v_F}) 2\log (1 + k_0/k) }
   {   k^2 + \frac{4 m }{\alpha v_F} +
           \frac{2N_f e^2}{\pi v_F} k^2 \log (1 + k_0/k)   }
\end{eqnarray}
We observe that the presence of the gap (accounted
for by the $m$ parameter) precludes an effective reduction
of the Coulomb interaction at small momenta. The dressed
potential $V_d (x)$ (for the same choice of parameters
made above) has been represented in the inset of
Fig. \ref{one}, where it is also compared with the bare
Coulomb potential.

From the above analysis, it is clear that the dressed interaction
keeps the long-range character of the Coulomb potential. Thus, in
the more complete picture accounting for the screening effects
within the nanotube, the result of having localized states due to
the presence of the fullerene charge remains unmodified. The
actual number of them depends on the value $Q_z$ of the net
fullerene charge, as well as on the length of the carbon nanotube.
The form of the spectrum within the gap can be obtained for a
particular system by diagonalizing the lattice hamiltonian made of
the tight-binding term (\ref{tb}) and the interaction term with
the dressed potential
\begin{equation}
H_{int} = \sum_{p,i} V_d (x_i - z) d^+_p (x_i) d_p (x_i)
\label{lat}
\end{equation}
where $x_i$ label the different sections
of atoms in the nanotube, in the same fashion as in
Eq. (\ref{tb}).

\begin{figure}
\begin{center}
\mbox{\epsfysize 4.5cm \epsfbox{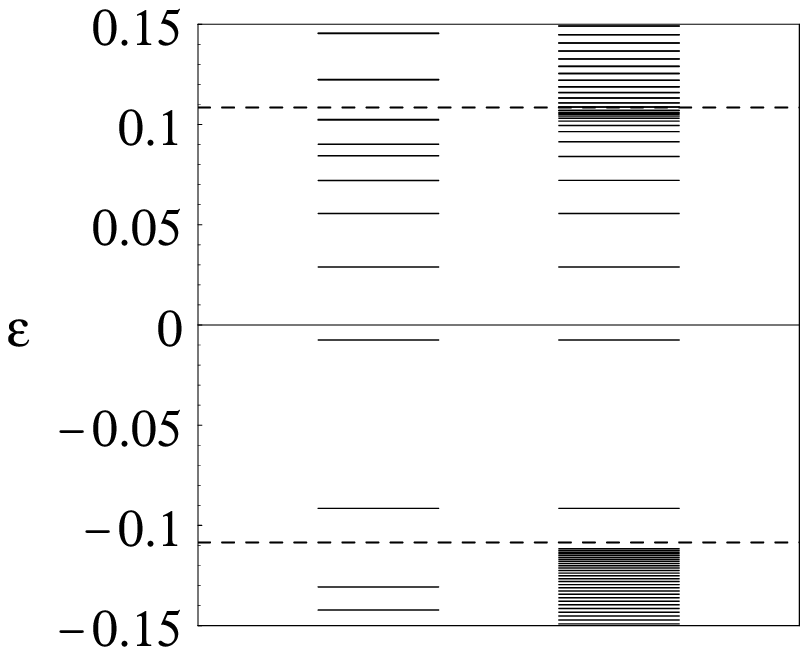}
\epsfysize 4.1cm 
\raisebox{0.2cm}{\epsfbox{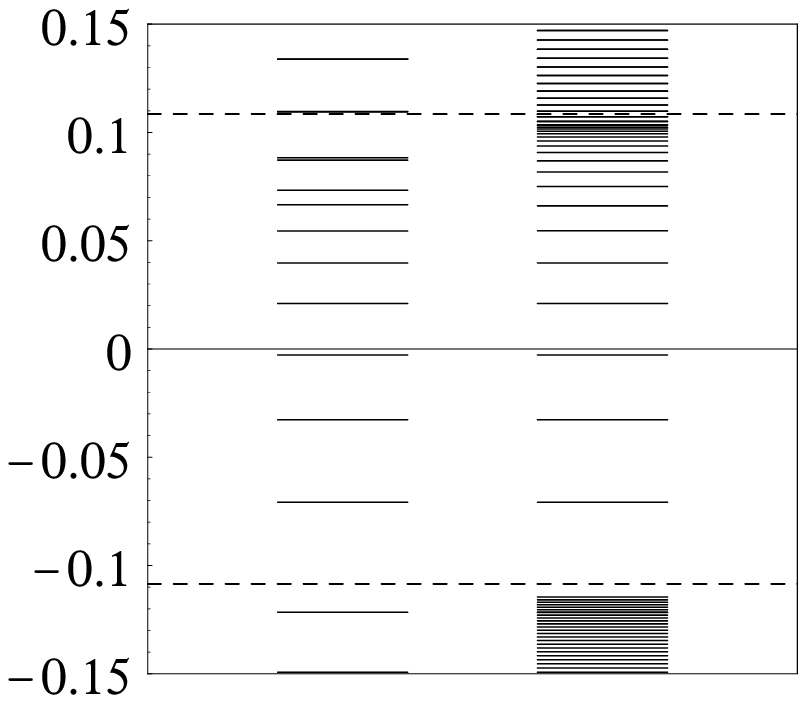}}}
\end{center}
\caption{Diagram of the low-energy levels obtained from
the diagonalization of the lattice hamiltonian for
$Q_z = 0.5$ (left) and $Q_z = 1.0$ (right). The spectra
are given in each case for a couple of nanotube segments
with lengths $L = 300a$ and $1500a$. The energies are
in units of the transfer integral $t$. The dashed lines
mark the border of the gap in the absence
of the fullerene charge.}
\label{two}
\end{figure}

The results of diagonalizing the hamiltonian $H_{tb} + H_{int}$
for a semiconducting (17,0) nanotube have been represented in Fig.
\ref{two}. The level of the deepest state within the gap depends
on the net charge $Q_z$, but there is always an accumulation of
states near the conduction band edge. Using the continuum
approximation, a state within the gap, at energy $- \Delta \le
\varepsilon < \Delta$, is localized within a length $l \approx v_F /
\sqrt{\Delta^2 - \varepsilon^2}$. This estimate agrees well with
the numerical calculations based on the tight-binding model.
Hence, the localization length of the deepest levels within the
gap is comparable to the radius of the nanotube. Note that each
subband will give rise to a set of localized states within its
gap.

\begin{figure}
\begin{center}
\vspace{0.5cm}
\mbox{\epsfysize 7.0cm
\epsfbox{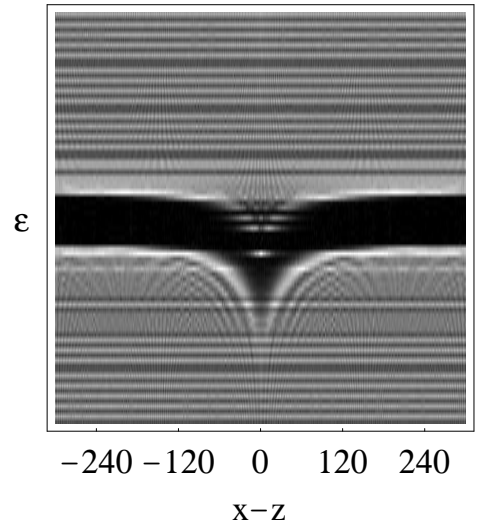}}
\end{center}
\caption{Intensity plot of the local density of states
obtained from the diagonalization of the lattice
hamiltonian for a (17,0) nanotube with $Q_z = 0.6$.
The variable $x$ is measured in units of the $C$-$C$
distance in the nanotube.}
\label{three}
\end{figure}

The existence of localized states within the gap of the
semiconducting nanotube has a significant impact on the
shape of the local density of states of the system. This
depends on the value of the net charge $Q_z$ and, for
each particular experimental setup, on the amount of
screening due to external charges. For the sake of
establishing a comparison with the spectra measured in
Ref. \onlinecite{lee}, we have adopted a phenomenological
approach by fixing the effective value of $Q_z$ so
that the deepest localized state is placed at the center
of the nanotube gap. The local density of states obtained
from the diagonalization of the lattice hamiltonian
$H_{tb} + H_{int}$ for
a (17,0) nanotube is represented in Fig. \ref{three}.
We observe the similarity with the local variation of the
gap measured in Ref. \onlinecite{lee} around the position
of each fullerene cluster. In
particular, we notice that the narrowing of the gap takes
place over a distance of a few nanometers,
in agreement with the spatial modulation of the gap
observed in the experiment.

Our results show that the Coulomb interaction between the
fullerene clusters and the nanotube has to be taken into
account to achieve a consistent explanation of the spatial
modulation of the gap in the peapods. In this respect, our
approach is complementary to those studies that have
focused on the hybridization of the molecular orbitals of
the fullerenes with the nanotube states. The present
analysis is consistent with the fact that the narrowing
of the gap has been measured in the peapods with
metallofullerenes, while no gap reduction has been observed
in the samples with encapsulated $C_{60}$ molecules.
Our investigation points at the possibility of having
control of the gap reduction in the peapods,
by varying the net fullerene charge $Q_z$. This
effect should be confronted in suitably prepared samples,
for the sake of making progress in the local bandgap
engineering of nanotube devices.


{\it Acknowledgements.} The financial support of the Ministerio 
de Educaci\'on y Ciencia (Spain) through grants
MAT2002-0495-C02-01 (for F. G.) and BFM2003-05317 (for J. G.)
is gratefully acknowledged.

\end{document}